\documentclass[journal]{IEEEtran}

\pdfoutput=1
\usepackage{tablefootnote}
\usepackage{stfloats}

\usepackage[flushleft]{threeparttable}
\usepackage[pdftex]{graphicx}
\graphicspath{{./Graficas}}
\usepackage{float}
\usepackage{graphicx,amsfonts,latexsym}
\usepackage{dsfont}
\usepackage{subcaption}
\usepackage{caption}
\usepackage{amsmath} 
\usepackage{amssymb}  
\usepackage{amsfonts}
\usepackage{amsthm} 
\usepackage{booktabs}

\newtheorem{definition}{Definition}

\newtheorem{property}{Property}
\newtheorem{remark}{Remark}

\newtheorem{corollary}{Corollary}

\newtheorem{proposition}{Proposition}

\usepackage{siunitx}
\usepackage{multirow}
\usepackage{arydshln}

\usepackage[ruled, lined, ,longend, linesnumbered]{algorithm2e}
\usepackage[hang,flushmargin]{footmisc}
\usepackage{algpseudocode}
\SetKwInOut{Input}{Input}
\SetKwInOut{Declares}{Declares}
\SetKwInOut{Require}{Require}
\SetKwInput{Prototype}{Prototype}
\usepackage{algpseudocode}

\usepackage{authblk}
\usepackage[T1]{fontenc}
\usepackage[utf8]{inputenc}

\usepackage{xcolor}
\usepackage{hyperref}
\hypersetup{%
   colorlinks=true,
   linkcolor={blue!50!black},
   citecolor={blue!50!black},
   urlcolor={blue!50!black}
}

\usepackage{fancyhdr}

\newcommand {\bsis} {\left\{ \begin{array} }
\newcommand {\esis} {\end{array}\right.}

\def\bmat#1{\left[\begin{array}{#1}}
	\def\emat{\end{array}\right]}

\def\diag{\texttt{diag}}

\def\({\left(}
\def\){\right)}

\def\vv#1{{ \rm \bf{#1}}}
\def\Id{{\rm{I}}}

\def\R{\mathbb{R}}    
\def\Z{\mathbb{Z}}
\def\fracg#1#2{{\displaystyle{\frac{#1}{#2}}}}  

\newcommand {\T}{^{\top}} 
\newcommand{\blista}{\renewcommand{\labelenumi}{(\roman{enumi})} 
	\begin{enumerate}}
	\newcommand{\elista}{\end{enumerate} \renewcommand{\labelenumi}{\arabic{enumi}.}}
\newcommand{\sCost}[1]{\beta_{#1}}

\def\LBx{\underline{x}}
\def\UBx{\overline{x}}
\def\LBu{\underline{u}}
\def\UBu{\overline{u}}
\def\LBy{\underline{y}}
\def\UBy{\overline{y}}

\def\LBv{\underline{v}}
\def\UBv{\overline{v}}

\makeatletter 
\newcommand\semiHuge{\@setfontsize\semiHuge{22.72}{27.38}}
\makeatother

\begin{document}
\pagestyle{fancy}
\title{\semiHuge Implementation of soft-constrained MPC for Tracking using its semi-banded problem structure}
\author{Victor~Gracia$^{\dagger}$,~Pablo~Krupa$^\star$,~Daniel~Limon$^\dagger$,~Teodoro~Alamo$^\dagger$%
\thanks{$^\dagger$ Dept. of Systems Eng. and Automation, Universidad de Sevilla, Spain.}%
\thanks{$^\star$ Gran Sasso Science Institute (GSSI), L'Aquila, Italy.}%
\thanks{The authors acknowledge support from grants PID2022-142946NA-I00 and PID2022-141159OB-I00 funded by MCIN/AEI/ 10.13039/501100011033 and by ERDF A way of making Europe; the MUR-PRO3 project on Software Quality; and the MUR-PRIN project DREAM (20228FT78M). Corresponding author: Victor Gracia. E-mails: \texttt{vgracia@us.es}, \texttt{pablo.krupa@gssi.it}, \texttt{dlm@us.es}, \texttt{talamo@us.es}}
}

\maketitle
\thispagestyle{fancy}

\begin{abstract}
	Model Predictive Control (MPC) is a popular control approach due to its ability to consider constraints, including input and state restrictions, while minimizing a cost function. However, in practice, these constraints can result in feasibility issues, either because the system model is not accurate or due to the existence of external disturbances. To mitigate this problem, a solution adopted by the MPC community is the use of soft constraints. In this article, we consider a not-so-typical methodology to encode soft constraints in a particular MPC formulation known as MPC for Tracking (MPCT), which has several advantages when compared to standard MPC formulations. The motivation behind the proposed encoding is to maintain the semi-banded structure of the ingredients of a recently proposed solver for the considered MPCT formulation, thus providing an efficient and fast solver when compared to alternative approaches from the literature. We show numerical results highlighting the benefits of the formulation and the computational efficiency of the solver.
\end{abstract}


\section{Introduction} \label{sec:introduction}

\IEEEPARstart{M}{odel} Predictive Control (MPC) \cite{rawlings2017model} is an advanced control methodology able to deal with (possibly joint) constraints in  inputs and states while optimizing a performance index. A subclass of MPC formulations is linear MPC, characterized by considering a linear prediction model and convex inequality constraints. In practical implementations, the underlying optimization problem is a Quadratic Program (QP), which generally can be solved efficiently \cite{boyd2004convex}. Linear MPC, therefore, is a sensible option in practice, given the limitations in computational power and memory of some control systems. 

A problem that MPC faces in real applications is that its optimization problem may be infeasible for the current system state due to the presence of the state/input constraints. Furthermore, even if the MPC problem is currently feasible, feasibility may be lost in the next sample time due to inaccuracies in the prediction model and/or external disturbances. This issue is particularly relevant in the case of linear MPC, where model mismatch is almost always present.
A practical approach to deal with this issue is the use of soft constraints, where the constraints are allowed to be violated at a certain cost. For instance, in \cite{melanie_LCSS_soft}, the authors use a classical soft-constrained approach to implement standard MPC, where non-negative slack variables are included in the constraints and penalized in the cost function.

Some feasibility problems from standard MPC formulations are addressed by the MPC for Tracking (MPCT) formulation \cite{LIMON20082382}, such as infeasibility when the reference is not reachable, or when it undergoes abrupt online changes. This formulation is characterized by including some extra decision variables that work as an artificial reference. This change implies a notably larger domain of attraction and feasibility region, where the drawback lies on a typically more complex QP~to~be solved. This complexity may prevent the use of MPCT in fast applications if the problem structure is not suitably addressed. In \cite{gracia2024_ECC}, an efficient solver for MPCT is proposed, providing computation times of the same order as the ones required to solve standard linear MPC using state-of-the-art QP solvers such as OSQP \cite{osqp} or SCS \cite{scs}, thanks to the exploitation of the semi-banded structure of the ingredients of the Alternating Direction Method of Multipliers (ADMM) \cite{MAL-016} when applied to the MPCT formulation. Another efficient solver for the same formulation was proposed in \cite{krupa2021implementation}, where an extended ADMM algorithm is used instead.

Even though MPCT typically suffers from less feasibility issues than standard linear MPC, its optimization problem may still become infeasible, for the same reasons previously stated. 
Thus, the use of soft constraints in MPCT is still a reasonable choice in practice. Indeed, in \cite{melanie_MPCT_soft}, the authors propose a soft-constrained MPCT formulation by means of the slack variables approach. 
In this article, we propose an alternative way of encoding the soft constraints, which allows us to retain the simple semi-banded structure exploited by the solver proposed in \cite{gracia2024_ECC}. This results in a particularly simple way of dealing with the soft constraints, allowing us to provide a structure-exploiting solver for the soft-constrained MPCT formulation.

This article is structured as follows. Section \ref{sec:mpct_mod_formulation} presents the MPCT formulation and our proposed approach to encode soft constraints. Section \ref{sec:ADMM_mpct_mod_formulation} shows how the modified MPCT formulation can be efficiently solved by means of the ADMM algorithm. Section \ref{sec:numerical_results} provides numerical results demonstrating the practical benefits of the proposed solver when compared to alternative approaches and solvers. Finally, the main results of the article are summarized in Section \ref{sec:conclusion}.

\subsubsection*{Notation}
The set of positive definite matrices of size $n \times n$ is denoted as $\mathcal{S}^n_{\succ}$.
The set of integers from $a$ to $b$, both included, is written as $\Z_a^b$.
The $j$-th component of a vector $x$ is referred to as $x_{(j)}$.
The Euclidean norm of a vector $x$ is defined as $\|x\|_2 = \sqrt{x\T x}$.
For a matrix $Q\in \mathcal{S}^n_{\succ}$, $\| x \|_{Q} \doteq \sqrt{x\T Q x}$.
The identity matrix of dimension $n$ is denoted as $\Id_{n}$, and the zero matrix of dimension $n \times m$ as $0_{n \times m}$, respectively, where the sub-index may be omitted if the dimension is evident.
A vector of ones in $\mathbb{R}^n$ is referred to as $\mathds{1}_n $.
Component-wise inequalities of two vectors $x$ and $y$ is written as $x \leq (\geq) \; y$.
The symbol $\otimes$ refers to Kronecker product.
The concatenation of vectors $x_1$ to $x_N$ conforming a column vector is denoted as $(x_{1}, \dots, x_{N})$.
We denote the block diagonal matrix formed by the concatenation of $A_1$ to $A_N$ (possibly of different dimensions) as $\diag(A_1, A_2, \dots, A_N)$.
Given scalars $a_1$ to~$a_n$, $\max(a_1, \dots, a_n)$ returns the maximum element of $a_1$ to $a_n$.

\section{Soft-constrained MPC for tracking} \label{sec:mpct_mod_formulation}

Consider a discrete-time system of the form
\begin{subequations}\label{pred_model}
\begin{align}
&x(t+1) = Ax(t) + Bu(t),\\
&y(t) = Cx(t)+Du(t),
\end{align}
\end{subequations}
where $x(t) \in \R^{n_x}$, $u(t) \in \R^{n_u}$ and $y(t) \in \R^{n_y}$ are the state, input and output vectors at sample time $t$, respectively.
Assume that the system is constrained as
\begin{equation} \label{box_limits}
		\LBx  \leq  x(t) \leq \UBx,\quad
		\LBu  \leq  u(t) \leq \UBu,\quad
        \LBy  \leq  y(t) \leq \UBy, 
\end{equation}
where $\LBx, \UBx \in \R^{n_x}$, $\LBu, \UBu \in \R^{n_u}$, $\LBy, \UBy \in \R^{n_{y}}$ are such that $\LBx < \UBx$, $\LBu< \UBu$ and $\LBy< \UBy$. The control objective is to stabilize the system at an equilibrium point $(x_r,u_r)$, namely the reference, while satisfying the constraints \eqref{box_limits}. The controller should steer the system to $(x_r,u_r)$, provided it is admissible. Otherwise, it should converge to its closest admissible equilibrium point. 

A particularly suitable MPC formulation to this end is \textit{MPC for tracking} (MPCT) \cite{LIMON20082382}. In particular, we focus on the following MPCT formulation with terminal equality constraint\begin{subequations}\label{MPCT_formulation}
	\begin{align}
		\min_{\substack{\vv{x,u}, \\x_s,u_s}} \; & V_o(x_s,u_s;x_r,u_r) + \sum_{i=0}^{N-1} l(x_i,u_i,x_s,u_s)   \\ 
		\textrm{s.t.} \; & x_{0} = x(t),  \label{initial_constraint}\\
		& x_{i+1} = Ax_{i} + Bu_{i}, \ i \in \Z_{0}^{N-2},\\
		& x_{s} = Ax_{N-1} + Bu_{N-1},\\
        & x_{s} = Ax_{s} + Bu_{s}, \label{eq_point_constraint}\\
        & \LBu_i \leq u_{i} \leq \UBu_i, \ i \in \Z_0^{N-1}, \label{ineq_u}\\ 
        & \LBx_i \leq x_{i} \leq \UBx_i, \ i \in \Z_1^{N-1}, \label{ineq_x}\\
        & \LBy_i \leq C x_{i} + D u_{i}\leq \UBy_i, \ i \in \Z_0^{N-1}, \label{ineq_y}\\
		& \LBx_s \leq x_{s} \leq \UBx_s, \ \LBu_s \leq u_{s} \leq \UBu_s, \label{ineq_xs_us} \\
        & \LBy_s \leq C x_s + D u_s \leq \UBy_s \label{ineq_ys},
	\end{align}
\end{subequations}
where the decision variables vectors \vv{x} = ($x_{0}$, $\dots$, $x_{N-1}$), \vv{u} = ($u_{0}$, $\dots$, $u_{N-1}$) are the sequence of states and inputs predicted along the prediction horizon $N$, respectively; $x(t)$ is the state of the system at current sample time $t$; $(x_{s},u_{s})$ are decision variables that work as an artificial reference that is required to be an admissible steady state of the system by means of constraints \eqref{eq_point_constraint}, \eqref{ineq_xs_us}, and \eqref{ineq_ys}, which are responsible of the numerous benefits of MPCT over classical MPC; $l(x_i,u_i,x_s,u_s)=\|x_{i}{-}x_{s}\|_{Q}^2 + \|u_{i}{-}u_{s}\|_{R}^2$ is the stage cost and $V_o(x_s,u_s;x_r,u_r) = \|x_{s}{-}x_{r}\|_{T}^2 + \|u_{s}{-}u_{r}\|_{S}^2$ is the offset cost, where $Q \in \mathcal{S}^{n_x}_{\succ}$, $R \in \mathcal{S}^{n_u}_{\succ}$, $T \in \mathcal{S}^{n_x}_{\succ}$ and $S \in \mathcal{S}^{n_u}_{\succ}$ are weight matrices that can be chosen so as to achieve a desired closed-loop behaviour \cite{STATHOPOULOS20142285}, or to ensure properties such as local optimality of the MPCT controller \cite{FERRAMOSCA20091975}; $\LBx_i, \UBx_i, \LBx_s, \UBx_s \in \R^{n_x}$, $\LBu_i, \UBu_i, \LBu_s, \UBu_s \in \R^{n_u}$ and $\LBy_i, \UBy_i, \LBy_s, \UBy_s \in \R^{n_{y}}$ are vectors that impose the box constraints limits, with $\LBx_i < \UBx_i$, $\LBu_i < \UBu_i$ and $\LBy_i < \UBy_i$, $\forall i$, and $\LBx_s < \UBx_s$, $\LBu_s < \UBu_s$, $\LBy_s < \UBy_s$. Note that the sub-index $i$ in \eqref{ineq_u}, \eqref{ineq_x} and \eqref{ineq_y} allows changes in the limits \eqref{box_limits} along $N$ to make \eqref{MPCT_formulation} more flexible, providing the possibility of implementing tube-based robust MPC \cite{Alvarado_JPC_RMPC_20}.

Formulation \eqref{MPCT_formulation} features key characteristics such as recursive feasibility when controlling the system under nominal conditions from a feasible initial state, even when abrupt changes occur in $(x_r,u_r)$, and guarantees convergence to its closest admissible equilibrium point of the system \cite{LIMON20082382}, where distance is measured according to the offset cost function $V_o(\cdot)$.
However, in real implementations, discrepancies between the prediction model \eqref{pred_model}, strong disturbances, or even inaccuracies in the solution provided by the solver applied to \eqref{MPCT_formulation}, may eventually lead to an infeasible problem. To mitigate this issue, we now present a soft-constrained version of \eqref{MPCT_formulation}. 

Let us start by showing the approach used to ``soften'' inequality constraints in \eqref{MPCT_formulation}.
Given the vectors $w,\underline{w}, \overline{w} \in \mathbb{R}^p$ that determine a hard inequality constraint $\underline{w} \leq w \leq \overline{w}$ in an optimization problem, we can use a ``soft version'' of it by replacing the constraint with a penalizing cost that measures the violation of the constraint.
To this end, we denote
\begin{equation*}
    \gamma_\beta(w;\overline{w},\underline{w}) \doteq \sum_{j=1}^{p} \beta_{(j)} \text{max}(w_{(j)}-\overline{w}_{(j)},\underline{w}_{(j)}-w_{(j)},0),
\end{equation*}
where the non-negative vector $\beta\in\R^p$ weights the penalization associated to the different components of $w$.
Let us define
\begin{equation*}
  v_t \doteq (y_0, x_1, u_1, y_1, \dots, x_{N-1}, u_{N-1}, y_{N-1}, x_s, u_s, y_s),
\end{equation*}
where $y_i=Cx_i+Du_i$ and $y_s = Cx_s+Du_s$, and
\begin{equation}
  v \doteq (x_0, u_0, v_t),\label{def:v}  
\end{equation}
of dimension $n_v = (N+1)(n_x+n_u+n_y)$. 
Defining
\begin{equation*}
    \begin{aligned}
        \overline{v}_t \doteq & (\overline{y}_0 , \overline{x}_1 , \overline{u}_1, \overline{y}_1, \dots, \overline{x}_{N-1}, \overline{u}_{N-1}, \overline{y}_{N-1}, \overline{x}_s, \overline{u}_s, \overline{y}_s), \\ 
        \underline{v}_t \doteq & (\underline{y}_0, \underline{x}_1, \underline{u}_1, \underline{y}_1, \dots, \underline{x}_{N-1}, \underline{u}_{N-1}, \underline{y}_{N-1}, \underline{x}_s, \underline{u}_s,\underline{y}_s),
    \end{aligned}
\end{equation*}
we have that the constraints in \eqref{MPCT_formulation} impose $\underline{v}_t \leq v_t \leq \overline{v}_t$.
Thus, we can transform \eqref{MPCT_formulation} into its soft-constrained variant
\begin{subequations}\label{MPCT_formulation_soft}
	\begin{align}
		\min_{\substack{\vv{x,u}, \\x_s,u_s}} \; &  V_o(x_s,u_s;x_r,u_r)
	    {+} \gamma_\beta(v_t ; \overline{v}_t, \underline{v}_t) {+} \sum_{i=0}^{N-1} l(x_i,u_i,x_s,u_s) \label{functional_MPCT_soft} \\
		\textrm{s.t.} \; & x_0 = x(t),\\
		& x_{i+1} = Ax_{i} + Bu_{i}, \ i \in \mathbb{Z}_{0}^{N-2},\\
		& x_{s} = Ax_{N-1} + Bu_{N-1},\\
		& x_{s} = Ax_{s} + Bu_{s},\\
        & \LBu_0 \leq u_0 \leq \UBu_0, \label{MPCT_formulation_soft:hard_u0}
	\end{align}
\end{subequations}
where $\gamma_\beta(v_t ; \overline{v}_t, \underline{v}_t)$ penalizes the violation of the constraints $\LBu_i \leq u_i \leq \UBu_i$, $i \in \mathbb{Z}_1^{N-1}$, as well as \eqref{ineq_x}-\eqref{ineq_ys}. Note that we do not relax the constraints $\LBu_0 \leq u_0 \leq \UBu_0$, given that, in practice, bounds in system inputs are typically given by the physical limits of the actuators.
The main advantage of this soft-constrained approach w.r.t. \eqref{MPCT_formulation} is that the underlying optimization problem \eqref{MPCT_formulation_soft} is always feasible if $N$ is larger than the controllability index of the system, a condition which is assumed to be satisfied.
Thus,~\eqref{MPCT_formulation_soft} always admits an optimal solution; a condition that will guarantee the asymptotic convergence of the solver presented in the following section.

\begin{remark} \label{remark:exact_penalization}
The concept of exact penalty function guarantees that, for sufficiently large values of $\beta$, the optimal solution of \eqref{MPCT_formulation_soft} is equal to the optimal solution of \eqref{MPCT_formulation} if the latter is feasible \cite[Theorem~14.3.1]{doi:https://doi.org/10.1002/9781118723203.ch14}, \cite{Kerrigan2000SoftCA}. The selection of $\beta$ should depend on the importance given to constraints violation. When problem \eqref{MPCT_formulation} is not feasible, the value of $\beta$ affects the numerical conditioning of \eqref{MPCT_formulation_soft} and thus the performance of the solver.
\end{remark}

\section{Efficiently applying ADMM to softened MPCT} \label{sec:ADMM_mpct_mod_formulation}

Due to the non-smoothness of \eqref{functional_MPCT_soft}, we may think that the proposed formulation leads to a challenging optimization problem. However, we now present an ADMM-based solver that leverages the structure of the ingredients of the ADMM method when applied to the MPCT formulation \eqref{MPCT_formulation_soft} and efficiently deals with the soft constraints.

\subsection{Alternating Direction Method of Multipliers}

 Consider the convex, closed and proper functions $f:\mathbb{R}^{n_z} \rightarrow (-\infty,\infty]$ and $g:\mathbb{R}^{n_v} \rightarrow (-\infty,\infty]$. Let $z \in \mathbb{R}^{n_z}$, $v \in \mathbb{R}^{n_v}$, $E \in \R^{n_\lambda \times n_z}$ and $F \in \R^{n_\lambda \times n_v}$.
 The optimization problem considered by the ADMM optimization method is
\begin{subequations}\label{eq:ADMM_problem}
	\begin{align}
		\min_{z,v} & \; f(z) + g(v) \\
		\textrm{s.t.} & \; Ez+Fv=0,
	\end{align}
\end{subequations}
with augmented Lagrangian $\mathcal{L}_\rho {\colon} \mathbb{R}^{n_z} {\times} \mathbb{R}^{n_v} {\times} \mathbb{R}^{n_\lambda} {\rightarrow} (-\infty,\infty]$
\begin{equation*}
	\mathcal{L}_\rho(z,v,\lambda) = f(z) + g(v) + \lambda\T (Ez+Fv) + \frac{\rho}{2}\|Ez+Fv\|_2^2,
\end{equation*}
where $\lambda \in \mathbb{R}^{n_\lambda}$ gathers dual variables and $\rho>0$ is the penalty parameter. We refer to a solution of (\ref{eq:ADMM_problem}) as ($z^*, v^*, \lambda^*$), which is assumed to exist.
Under this assumption, ADMM (Algorithm~\ref{ADMM_algorithm}) asymptotically converges to the optimal solution of \eqref{eq:ADMM_problem} (see \cite{ADMM_convergence} for a convergence analysis). The residuals $\epsilon_p,\epsilon_{d}>0$, which must be selected on a cases-by-case basis, serve as exit conditions in practice to ensure a certain suboptimality level of the output ($\tilde{z}^*$, $\tilde{v}^*$, $\tilde{\lambda}^*$) of Algorithm~\ref{ADMM_algorithm}, as discussed in~\cite{MAL-016}.

\begin{algorithm}[t]
	\DontPrintSemicolon
	\caption{ADMM} \label{ADMM_algorithm}
	\Require{$v^{0}$, $\lambda^{0}$, $\rho>0$, $\epsilon_{p}>0$, $\epsilon_{d}>0$}
	$k \gets 0$\;
	\Repeat{{$\|Ez^{k}+Fv^{k}\|_{\infty} \leq \epsilon_{p}$ and $\|v^{k}-v^{k-1}\|_{\infty} \leq \epsilon_{d}$}}{
		$z^{k+1} \gets \displaystyle \arg \min_{z} \mathcal{L}_{\rho}(z, v^{k}, \lambda^{k})$\; \label{step_z_ADMM}
		$v^{k+1} \gets \displaystyle \arg \min_{v} \mathcal{L}_{\rho}(z^{k+1}, v, \lambda^{k})$\; \label{step_v_ADMM}
		$\lambda^{k+1} \gets \lambda^{k} + \rho(Ez^{k+1} + Fv^{k+1})$\;
		$k \gets k+1$\;
	}		
	\KwOut{$\tilde{z}^{*} \gets z^{k}$, $\tilde{v}^{*} \gets v^{k}$, $\tilde{\lambda}^{*} \gets \lambda^{k}$} 
\end{algorithm}

\subsection{Applying ADMM to MPCT}

To solve \eqref{MPCT_formulation_soft} using Algorithm \ref{ADMM_algorithm}, we first transform \eqref{MPCT_formulation_soft} into \eqref{eq:ADMM_problem}.
Taking $v$ defined in \eqref{def:v} and 
\begin{equation*}
    z \doteq (\tilde{x}_0, \tilde{u}_0, \tilde{x}_1, \tilde{u}_1, \dots, \tilde{x}_{N-1}, \tilde{u}_{N-1}, \tilde{x}_s, \tilde{u}_s),
\end{equation*}
where $\tilde{x}_s, \tilde{x}_i \in \R^{n_x}$, $\tilde{u}_s, \tilde{u}_i \in \R^{n_u}$, $i \in \Z_0^{N-1}$, and defining the indicator function $\mathcal{I}_{(G\hat{z}=b)}(\hat{z})= 0$ if $G\hat{z}=b$ and $\mathcal{I}_{(G\hat{z}=b)}(\hat{z})= +\infty$ otherwise,
we have that
\begin{equation}
		f(z) = \frac{1}{2} z\T H z + q\T z + \mathcal{I}_{(Gz=b)}(z), \label{eq:ADMM:MPCT:f} 
\end{equation}
where $q = -(0, 0, \dots, 0, Tx_r, Su_r)$, $b = (x(t) , \dots, 0)$,
\begin{subequations} \label{eq:ADMM:MPCT:ingredients}
\begin{align} 
		&H =\begin{bmatrix}
				Q & 0 & \cdots & -Q & 0\\
				0 & R & \cdots & 0 & -R\\
				0 & 0 & \ddots & \vdots & \vdots\\
				-Q & 0 & \cdots & NQ+T & 0\\
				0 & -R & \cdots & 0 & NR+S
			\end{bmatrix}, \label{eq:ADMM:MPCT:ingredients:H} \\
 		&G =\begin{bmatrix}
 				\Id & 0 & 0 & 0 & \cdots & 0\\
 				A & B & -\Id & 0 & \cdots &  0\\
 				0 & \ddots & \ddots & \ddots & 0 & \vdots\\
 				0 & 0 & A & B & -\Id & 0\\
 				0 & 0 & 0 & 0 & (A-\Id) & B
 			\end{bmatrix}, \label{eq:ADMM:MPCT:ingredients:G}
 \end{align}
\end{subequations}
with $G \in \R^{m_z\times n_z}$, $n_z = (N+1)(n_x+n_u)$, $m_z = (N+2)n_x$.
Defining now, for $a,\overline{a},\underline{a}\in\R^{n_a}$, the indicator function $\mathcal{I}_{[\underline{a},\overline{a}]}(a)= 0$ if $\underline{a} \leq a \leq \overline{a}$ and $\mathcal{I}_{[\underline{a},\overline{a}]}(a)= +\infty$ otherwise, 
we have that
    $g(v) = \gamma_{\beta/2}(v_t ; \overline{v}_t , \underline{v}_t) + \mathcal{I}_{[\underline{u}_0,\overline{u}_0]}(u_0)$.
Now, let
$$\hat{E} = 
        \begin{bmatrix}
            \Id_{n_x} & 0 \\
             0 & \Id_{n_u} \\
             C & D
        \end{bmatrix}.$$
Then, we have $E = \diag(\hat{E}, \dots, \hat{E})$, $F = -\Id_{n_v}$, $n_\lambda = n_v$.

\subsection{Computation of Step~\ref{step_z_ADMM} of Algorithm~\ref{ADMM_algorithm}}

Step~\ref{step_z_ADMM} in Algorithm~\ref{ADMM_algorithm} provides $z^{k+1}$, which consists in minimizing $\mathcal{L}_\rho(z,v^k,\lambda^k)$ over $z$. Particularizing it for \eqref{eq:ADMM:MPCT:f}, $E$ and $F$, we have
\begin{subequations} \label{z_problem}
    \begin{align}
z^{k+1} = \arg \min_{z} & \; \frac{1}{2} z\T P z + (p^k)\T z \\
\textrm{s.t.} & \; Gz=b,
    \end{align}
\end{subequations}
where $P = H + \rho E\T E$ and $p^k = q + E\T(\lambda^k - \rho v^k)$. Matrix $P$ is "semi-banded", a term we now formally introduce.

\begin{definition}
    A non-singular matrix $M \in \R^{n \times n}$ is semi-banded if it can be decomposed as $\Gamma + UV$, where $\Gamma \in \R^{n \times n}$ is banded diagonal and full rank, $U \in \R^{n \times m}$, $V \in \R^{m \times n}$, with $n \gg m$. We say a linear system $Mz=d$ is semi-banded if $M$ is semi-banded.
\end{definition}
The authors in \cite{gracia2024_ECC} propose a way of efficiently solving \eqref{z_problem} by decoupling the semi-banded structure of the problem, which reduces the computational cost compared to generic QP solvers. We provide here a brief description of the approach for completeness and refer the reader to \cite{gracia2024_ECC} for further details.

\begin{proposition}[{\cite[\S 5.5.3]{boyd2004convex}}] 
    A sufficient and necessary condition for $z^*$ to be an optimal solution of \eqref{z_problem} is the existence of a vector $\mu \in \R^{m_z}$ satisfying
    \begin{equation} \label{eq:ecQP_KKT_conditions}
        Gz^*=b,\ Pz^*+G\T\mu+p^k = 0.
    \end{equation}
\end{proposition}

Defining $W \doteq GP^{-1}G\T$, we can express \eqref{eq:ecQP_KKT_conditions} as
\begin{subequations}\label{eq:three_steps_ecQP_KKT_conditions}
    \begin{align}
        & P \xi = p^k, \label{eq:three_steps_1}\\
        & W \mu = -(G\xi + b), \label{eq:three_steps_2}\\
        & Pz^* = -(G\T \mu + p^k), \label{eq:three_steps_3}
    \end{align}
\end{subequations}
where $W \in \R^{m_z\times m_z}$ and $\xi \in \R^{n_z}$. The three linear systems \eqref{eq:three_steps_ecQP_KKT_conditions} can be solved efficiently to obtain $z^{k+1}$ thanks to the semi-banded structures of $P$ and $W$, combined with the following Woodbury matrix identity \cite{woodbury}.

\begin{property}[Woodbury matrix identity]\label{property:woodbury}
    Consider a semi-banded matrix $M$. If $\Id+V\Gamma^{-1}U$ is non-singular, then
    \begin{equation*}
        M^{-1} = (\Gamma+UV)^{-1} = \Gamma^{-1}-\Gamma^{-1}U(\Id+V\Gamma^{-1}U)^{-1}V\Gamma^{-1}. 
    \end{equation*}
\end{property}

\begin{proposition}[{\cite[Proposition~2]{gracia2024_ECC}}]\label{prop:P_W_semibanded}
    Denote $Y \doteq -\mathds{1}_{N}\T \otimes \diag(Q, R)$. Then, $P$ and $W$ in \eqref{eq:three_steps_ecQP_KKT_conditions} can be decomposed as $P=\widehat{\Gamma} + \widehat{U} \widehat{V}$ and $W=\tilde{\Gamma} + \tilde{U} \tilde{V}$ by taking
    \begin{equation*}
        \begin{aligned}
            &\widehat{\Gamma} = \diag(Q,R,Q,R,\dots,NQ+T,NR+S)+\rho E\T E,\\
            &\widehat{U} = \begin{bmatrix}
                Y\T & 0 \\
                0 & \Id_{(n_x+n_u)}
            \end{bmatrix},
            \widehat{V} = \begin{bmatrix}
                0 & \Id_{(n_x+n_u)}\\
                Y & 0
            \end{bmatrix}, \tilde{\Gamma} = G\widehat{\Gamma}^{-1}G\T,\\
            &\tilde{U} = -G\widehat{\Gamma}^{-1}\widehat{U}(\Id+\widehat{V}\widehat{\Gamma}^{-1}\widehat{U})^{-1}, \tilde{V}=\widehat{V}\widehat{\Gamma}^{-1}G\T.
        \end{aligned}
    \end{equation*}
    Furthermore, since matrix $G$ provided in \eqref{eq:ADMM:MPCT:ingredients:G} is full-column rank, matrices $P$, $W$, $\widehat{\Gamma}$ and $\tilde{\Gamma}$ are positive definite.
\end{proposition}

\begin{proposition}[{\cite[Proposition~3]{gracia2024_ECC}}]\label{prop:solve_with_alg_2}
    Consider the semi-banded system of equations $Mz=d$, where $M=\Gamma+UV$ satisfies Property~\ref{property:woodbury}. Then, Algorithm \ref{alg:solve_semi_banded} provides $\tilde{z}$ satisfying $M\tilde{z}=d$.
\end{proposition}

We refer the reader to \cite{gracia2024_ECC} for the proofs of Propositions~\ref{prop:P_W_semibanded} and \ref{prop:solve_with_alg_2}. Using the decompositions of $P$ and $W$ from Proposition~\ref{prop:P_W_semibanded}, Algorithm \ref{alg:solve_semi_banded} can be applied to obtain the solutions of \eqref{eq:three_steps_1}, \eqref{eq:three_steps_2} and \eqref{eq:three_steps_3}. Note that $z^*$ in \eqref{eq:three_steps_3} yields $z^{k+1}$.

\begin{remark}
    The computational cost of Steps \ref{alg:SB_step_1} and \ref{alg:SB_step_3} of Algorithm~\ref{alg:solve_semi_banded} when applied to equations \eqref{eq:three_steps_1} and \eqref{eq:three_steps_3} is low, as $\widehat{\Gamma}$ is block diagonal. Conversely, the same steps applied to \eqref{eq:three_steps_2} have a higher cost, since $\tilde{\Gamma}$ is banded-diagonal. As in \cite{gracia2024_ECC}, a Cholesky decomposition of $\tilde{\Gamma}$ can be used to efficiently solve Steps \ref{alg:SB_step_1} and \ref{alg:SB_step_3} of Algorithm \ref{alg:solve_semi_banded} when applied to \eqref{eq:three_steps_2}. Thus, the computational complexity of solving \eqref{z_problem} grows linearly in $N$~\cite{Krupa_TCST_20}, instead of the $\mathcal{O}(N^2)$ complexity obtained when using a non-sparse approach. 
    Solving Step~\ref{alg:SB_step_2} of Algorithm \ref{alg:solve_semi_banded} when applied to the three linear systems in \eqref{eq:three_steps_ecQP_KKT_conditions} is inexpensive given that $\widehat{U}$, $\widehat{V}$, $\tilde{U}$ and $\tilde{V}$ shown in Proposition \ref{prop:P_W_semibanded} lead to small-dimensional matrices $\Id + \widehat{V} \widehat{\Gamma}^{-1} \widehat{U}$ and $\Id + \tilde{V} \tilde{\Gamma}^{-1} \tilde{U}$.
    We note that all the aforementioned matrix inversions and factorizations are performed offline and stored in memory for their online use, leading to an online linear computational complexity of Algorithm \ref{alg:solve_semi_banded} with respect to $N$.
\end{remark}

\begin{algorithm}[t]
	\DontPrintSemicolon
	\caption{Solve semi-banded system  $(\Gamma+UV)\tilde{z}{=}d$ (From \cite[Algorithm~2]{gracia2024_ECC})} \label{alg:solve_semi_banded}
	\Require{$\Gamma$, $U$, $V$, $d$}
    Compute $z_1$ solving $\Gamma z_1 = d$\; \label{alg:SB_step_1}
    Compute $z_2$ solving $(\Id + V \Gamma^{-1} U) z_2 = V z_1$\; \label{alg:SB_step_2}
    Compute $z_3$ solving $\Gamma z_3 = U z_2$\; \label{alg:SB_step_3}
    \KwOut{$\tilde{z} \gets z_1-z_3$}
\end{algorithm}

\subsection{Efficient computation of Step~\ref{step_v_ADMM} of Algorithm~\ref{ADMM_algorithm}}

The iterate $v^{k+1}$ is obtained in Step \ref{step_v_ADMM} of Algorithm \ref{ADMM_algorithm} as
\begin{equation}\label{QP_ineq_soft}
    \begin{aligned}
  v^{k+1} = \arg \min_{x_0,u_0,v_t} &\; \gamma_{\beta/2}(v_t ; \overline{v}_t , \underline{v}_t) - \lambda\T v + \frac{\rho}{2} \| Ez-v \|_2^2 \\
        \textrm{s.t.} &\; \underline{u}_0 \leq u_0 \leq \overline{u}_0,
    \end{aligned}
\end{equation}
which is separable. Note that it considers a non-differentiable piece-wise quadratic functional due to the inclusion of the $\max(\cdot)$ operator in the $\gamma_{\beta/2}(\cdot)$ term. Since \eqref{QP_ineq_soft} is separable, each element of its optimal solution can be obtained from a scalar optimization problem, whose solution is provided by the following proposition. We provide its proof in the Appendix.

\begin{proposition}\label{Prop: Max_Value_Problem}
Consider the strictly convex scalar problem
\begin{equation} \label{QP_max}
    y^*= \arg\min_{y\in \R} \ \frac{1}{2} y^2 - b y + \alpha \max (c-y,y-d,0),
\end{equation}
where $\alpha >0$, $b,c,d \in \mathbb{R}$ and $c<d$.
Define $y_1 = b+\alpha$, $y_2 = b$, $y_3 = b-\alpha$.
Then, \begin{equation*}
    y^*=
    \begin{cases}
        y_1 & \text{if} \ y_1 \leq c, \\
c & \text{if} \ y_1 > c  \ \text{and} \ y_2 < c, \\	
        y_2 & \text{if} \ c \leq y_2 \leq d, \\
d & \text{if} \ y_2 > d\ \text{and} \ y_3 < d  , \\ 				
y_3 & \text{if} \ y_3 \geq d. \\	 
    \end{cases}
\end{equation*}
 \end{proposition}

Proposition~\ref{Prop: Max_Value_Problem} leads to the following corollary, which provides the optimal solution for the elements $v_t$ of \eqref{QP_ineq_soft}. We provide no proof because the result is derived by simply equating terms between \eqref{QP_ineq_soft} and \eqref{QP_max}. 

\begin{corollary} \label{corollary}
    Denote $c_{(j)} \doteq (Ez^{k+1}+ \frac{1}{\rho} \lambda^{k})_{(j+n_x+n_u)}$, for $j\in\Z_{1}^{n_v-n_x-n_u}$.
    Then, each component of $v_t^{k+1}$ is given by
    \begin{equation*}
    v_{t(j)}^{k+1} =
    \begin{cases}
        c_{(j)} + \fracg{\sCost{(j)}}{2\rho} & \text{if} \ c_{(j)} + \fracg{\sCost{(j)}}{2\rho} < \LBv_{t(j)}, \\
        \LBv_{t(j)} & \text{if} \ c_{(j)} + \fracg{\sCost{(j)}}{2\rho} > \LBv_{t(j)} \ \text{and} \ c_{(j)} < \LBv_{t(j)}, \\
        c_{(j)} & \text{if} \ \LBv_{t(j)} \leq c_{(j)} \leq \UBv_{t(j)}, \\
        \UBv_{t(j)} & \text{if} \ c_{(j)} > \UBv_{t(j)} \ \text{and} \ c_{(j)} - \fracg{\sCost{(j)}}{2\rho} < \UBv_{t(j)}, \\
        c_{(j)} - \fracg{\sCost{(j)}}{2\rho} & \text{if} \ c_{(j)} - \fracg{\sCost{(j)}}{2\rho} \geq \UBv_{t(j)}.
    \end{cases}
\end{equation*}
\end{corollary}

Now, let
$\overline{v} \doteq (\mathds{1}_{n_x} \cdot \infty,\overline{u}_0), \ \underline{v} \doteq (-\mathds{1}_{n_x} \cdot \infty,\underline{u}_0)$. Then, the rest of elements of $v^{k+1}$, which are the ones corresponding to $(x_0,u_0)$, i.e., $v_{(j)}^{k+1}$, $j \in \mathbb{Z}_{1}^{n_x+n_u}$, can be computed as 
\begin{equation*}
    v_{(j)}^{k+1} = \min \left( \max \left( Ez_{(j)}^{k+1} + \frac{1}{\rho} \lambda_{(j)}^k,\underline{v}_{(j)} \right),\overline{v}_{(j)} \right).
\end{equation*}

\begin{figure}
    \centering
    \captionsetup{belowskip=-10pt}
    \includegraphics[height=1.55cm,width=0.4\textwidth]{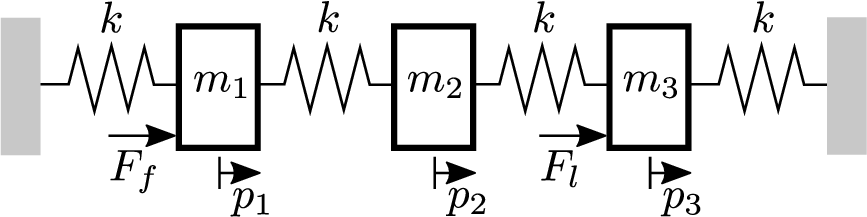}
    \caption{ Oscillating masses system}
    \label{fig: system}
\end{figure}

{\renewcommand{\arraystretch}{1.0}
\begin{table*}[t]
    \centering
    \setlength{\tabcolsep}{8pt}
    \captionsetup{belowskip=-15pt}
    \begin{tabular}{c|c|rrrr|rrrr}
    	\multirow{2}{*}{Formulation} & \multirow{2}{*}{Solver} & \multicolumn{4}{c|}{Iterations} &   \multicolumn{4}{c}{Computation time [milliseconds]}  \\ \cline{3-10}   
         &  & Avg.         & Median        & Max.	    & Min.   & Avg. & Median & Max. & Min.
        \\ \hline 
         & SPCIES &  30.7          & 31.0            & 45.0	             & 27.0   & 0.38 & 0.37 & 0.69 & 0.32 \\
        \eqref{MPCT_formulation_soft} & OSQP & 76.0           & 75.0            & 175.0	             & 50.0   & 1.49 & 1.42 & 3.30 & 1.08 \\
		 & SCS &  481.1         &    450.0        &  1350.0            & 250.0   & 12.50 & 12.40 & 35.00 & 6.78 \\ \hdashline[1pt/2pt]
         & SPCIES & 30.7           & 31.0            & 45.0	             & 27.0   & 0.37 & 0.36 & 0.61 & 0.32 \\
        \eqref{MPCT_formulation} & OSQP & 46.8           & 50.0            & 75.0	             & 25.0   & 0.62 & 0.62 & 1.03 & 0.42 \\
         & SCS &  417.5          &     400.0        &    750.0          & 300.0  & 22.70 & 22.30 & 44.80 & 16.10 \\ \hdashline[1pt/2pt]
       \multirow{2}{*}{\cite{melanie_MPCT_soft}$^*$} & OSQP & 50.1           & 50.0            & 75.0	             & 50.0   & 2.40 & 2.32 & 4.70 & 2.28 \\
         & SCS &  547.8          &  500.0           & 	1125.0             &  250.0  & 29.00 & 27.00 & 59.80 & 13.50 \\
        \hline
    \end{tabular}\\
    \footnotesize{*The soft-constrained formulation from \cite{melanie_MPCT_soft} has been implemented using the quadratic offset cost from this article for a fair comparison.}
\caption{Number of iterations and computation times.}
\label{table:time_iterate_comparison}
\end{table*}}

\section{Numerical results} \label{sec:numerical_results}

Let us consider a system, inspired by the case study in \cite{KOGEL20111362}, consisting of three masses connected by springs, with the ones to the far left and right connected to unmovable walls, as illustrated in Figure~\ref{fig: system}. The inputs are the two forces applied to each of the masses connected to the walls, i.e., $F_f$ and $F_l$, whereas the state vector is given by $x=(p_1, p_2, p_3, v_1, v_2, v_3)$, where $p_i$ and $v_i$ are the position and velocity of mass $i$, respectively. The physical parameters of the system are $m_1=m_2=m_3 = 1$kg, $k = 2$N/m.
A model \eqref{pred_model} is obtained by discretizing the continuous dynamics of the system using a sample time of $0.2$ seconds.
We consider two outputs, which are the relative distance between the center of masses $m_1$ and $m_2$, and the center of masses $m_2$ and $m_3$. Therefore, we have 
\begin{equation*}
 C =
    \begin{bmatrix}
        -1 & 1 & 0 & 0 & 0 & 0\\
        0 & -1 & 1 & 0 & 0 & 0
    \end{bmatrix}, \,
D =
    \begin{bmatrix}
        0 & 0 \\
        0 & 0
    \end{bmatrix}.
\end{equation*}

The constraints we consider for the system are
\begin{equation} \label{numerical: bounds}
    \begin{aligned}
        & \overline{x} = (0.6,0.6,0.6,1,1,1), \; \underline{x} = -\overline{x}, \; \overline{x}_s = \overline{x}, \; \underline{x}_s = \underline{x},\\
        & \overline{u} = (1,1), \; \underline{u} = (0,0), \; \overline{u}_s = \overline{u}, \; \underline{u}_s  = \underline{u}.
    \end{aligned}
\end{equation}
All the bounds in \eqref{MPCT_formulation_soft} are set to \eqref{numerical: bounds}.
The control objective is to steer the system from an initial state to the admissible equilibrium point $x_r = (0.4,0.4,0.4,0,0,0)$, $u_r = (0.8,0.8)$.

We set the parameters of \eqref{MPCT_formulation_soft} to $\beta = 20 \cdot \mathds{1}_{n_v}$, $N = 15$, $Q = \texttt{diag}(2.5,2.5,2.5,0.5,0.5,0.5)$, $R = \texttt{diag}(0.3,0.3)$, $S = \texttt{diag}(1,1)$, $T = \texttt{diag}(200,200,200,10,10,10)$,
and select $\rho = 1.2$ and $\epsilon_p=\epsilon_d=10^{-4}$ for ADMM.
We set the same parameters for \eqref{MPCT_formulation}. We perform a set of 1000 experiments, using an i5-1135G7 processor, where $x(t)$ are randomly selected from a uniform distribution within the range $-(0.1,0.1,0.1,0.2,0.2,0.2) \leq x(t) \leq (0.1,0.1,0.1,0.2,0.2,0.2)$, selected so as to ensure that all experiments are feasible for \eqref{MPCT_formulation}, but with a significant number of them leading to active constraints in their optimal solution.  

The results are shown in Table \ref{table:time_iterate_comparison}, where we implement~\eqref{MPCT_formulation} and \eqref{MPCT_formulation_soft} using version \texttt{v0.3.11} of the SPCIES toolbox for MATLAB \cite{Spcies}.
We also solve them using OSQP \cite{osqp} (version \texttt{0.6.2}) and SCS \cite{scs} (version \texttt{3.2.3}), providing a comparison with state-of-the-art operator-splitting QP solvers.
Additionally, using OSQP and SCS, we include the soft-constrained MPCT formulation from~\cite{melanie_MPCT_soft} (but using the $V_o(\cdot)$ of \eqref{MPCT_formulation} for fair comparison), to compare the classical slack-variable approach to implement soft constraints with the approach of Section~\ref{sec:mpct_mod_formulation}.
We use the 1-norm for penalizing the slack variables of~\cite{melanie_MPCT_soft}, and take $\xi=0.001$, $S = 20 \cdot \Id$, $\rho_\epsilon = 10$.

The results indicate that the proposed structure-exploiting solver for \eqref{MPCT_formulation_soft} provides better computational results than when using general-purpose QP solvers. Note also that, using our approach,  the computation time per iteration is only a bit larger for \eqref{MPCT_formulation_soft} than for \eqref{MPCT_formulation}, given the extra operations described in Corollary \ref{corollary}. Additionally, regarding the OSQP and SCS results, we notice that the slack-variables approach from \cite{melanie_MPCT_soft} is less efficient than the one we propose, mainly due to the inclusion of the slack variables as additional decision variables of the optimization problem, which increase its complexity and lead to the loss of its simple semi-banded structure.

\begin{figure*}[t]
    \centering
    \begin{subfigure}{0.33\textwidth}
        \includegraphics[width=\linewidth, height = 2.9cm]{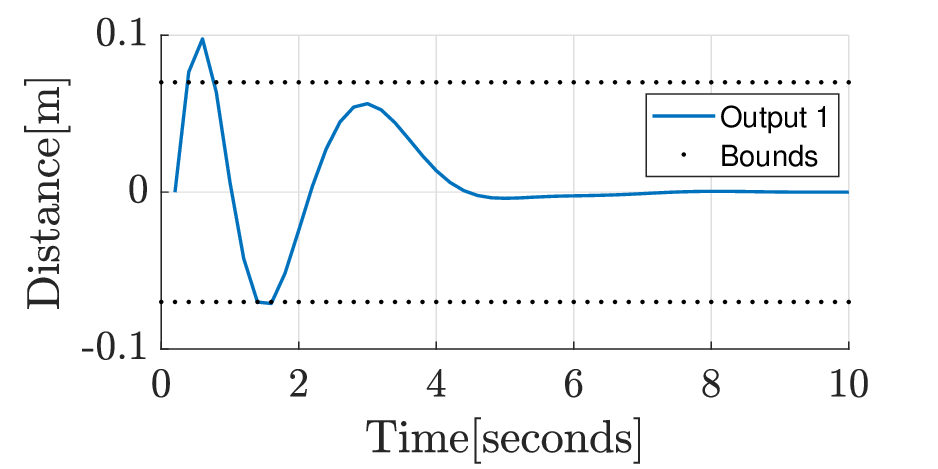}
        \caption{System output 1.}
    \end{subfigure}%
    \hfill
    \begin{subfigure}{0.33\textwidth}
        \includegraphics[width=\linewidth, height = 2.9cm]{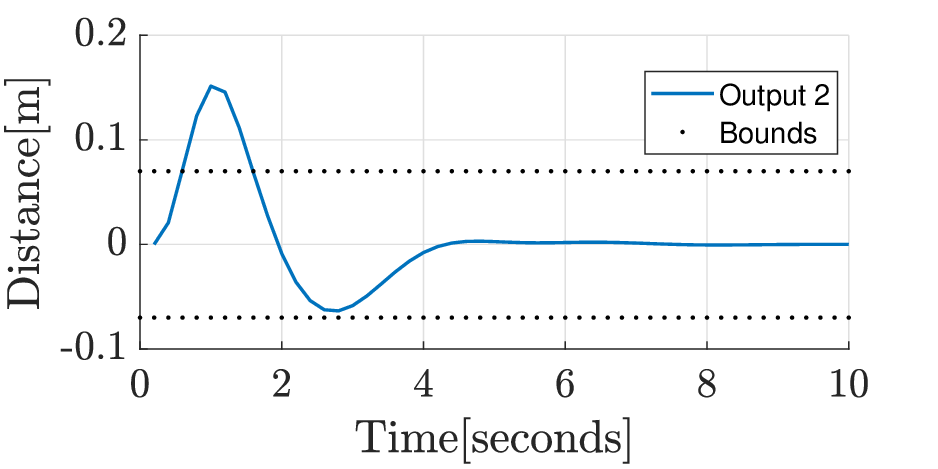}
        \caption{System output 2.}
    \end{subfigure}%
    \begin{subfigure}{0.33\textwidth}
        \includegraphics[width=\linewidth, height = 2.9cm]{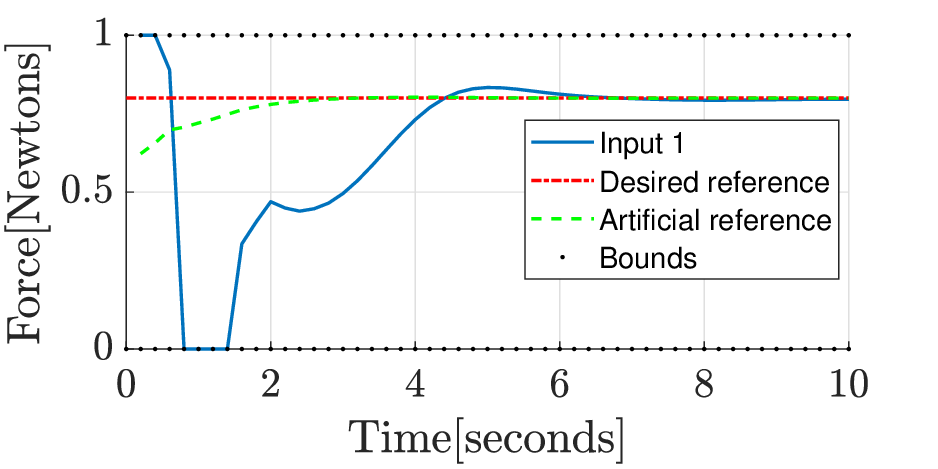}
        \caption{System input 1.}
    \end{subfigure}%
    \caption{\small Closed-loop experiment of the oscillating masses system using the soft-constrained formulation \eqref{MPCT_formulation_soft}.}
    \label{fig: closed-loop experiment}
\end{figure*}

\begin{remark}
    A relevant difference between \eqref{MPCT_formulation_soft} and the soft-constrained MPCT formulation in \cite{melanie_MPCT_soft} is that, in the former, only the first input of the predicted sequence along $N$ is subject to hard constraints, whereas, in the latter, the entire sequence of inputs is subject to hard constraints. Consequently, in contrast to \eqref{MPCT_formulation_soft}, the formulation in \cite{melanie_MPCT_soft} is not always~feasible.
\end{remark}
Next, we include closed-loop results starting from an initial state $x(t)=(0,0,0,-0.5,0,0,0,0)$, considering output limits $\overline{y} = (0.07,0.07)$, $\underline{y} = -\overline{y}$, $\overline{y}_s = \overline{y}$, $\underline{y}_s  = \underline{y}$,
while using $\rho=20$ and maintaining the rest of parameters the same. In this case, \eqref{MPCT_formulation} is infeasible, unlike \eqref{MPCT_formulation_soft}.
Results are presented in Figure~\ref{fig: closed-loop experiment}, showing that the applied input does not violate its constraints, as \eqref{MPCT_formulation_soft:hard_u0} considers hard inequality constraints for $u_0$. We remark that, when \eqref{MPCT_formulation} is not feasible, the iterations required by \eqref{MPCT_formulation_soft} depend on $\beta$. In this experiment, we observe 271 iterations on average, and a maximum of 506.

\section{Conclusion} \label{sec:conclusion}
This article has introduced an efficient way of implementing the MPCT formulation with most of its hard inequality constraints softened, so that the resulting optimization problem is always feasible.
This provides the ability to deal with situations where model mismatch or disturbances might cause the original formulation to become infeasible.
We propose an encoding of the soft constraints that allows us to retain the semi-banded structure of the formulation that is exploited by a recently proposed ADMM-based solver. Numerical results show that the proposed approach provides good computational results when compared to state-of-the-art QP solvers and with the classical slack-variable approach to encode soft constraints. Future studies will delve into the suitable online selection of the ADMM penalty parameter $\rho$, numerical preconditioning and warm-start procedures to improve performance.
\bibliographystyle{IEEEtran}
\bibliography{IEEEabrv,Biblio_ADMM_MPC}

\begin{appendices}\label{sec:appendix}
\renewcommand{\thesectiondis}[2]{:}

\section{Proof of Proposition \ref{Prop: Max_Value_Problem}} \label{app:proof:max_value}

    Clearly, the function to be minimized 
    $$ h(y)= \frac{1}{2} y^2 - b y + \alpha \max (c-y,y-d,0),$$
    can be written as $h(y)= \max\{h_1(y),h_2(y), h_3(y)\}$, where  
    \begin{eqnarray*} h_1(y) &=& \frac{1}{2}y^2 -by + \alpha(c-y), \ h_2(y) = \frac{1}{2}y^2 -by, \\
    h_3(y) &=& \frac{1}{2}y^2 -by + \alpha(y-d). 
    \end{eqnarray*}
    We can check that the proposed values for $y_1$, $y_2$ and $y_3$ satisfy 
    $$  y_i  =  \arg\min_{y\in \R} h_i(y), \; i=1,2,3. $$
    Suppose now that $y_1 \leq c$, and recall that $c < d$. Then
    $$ h(y_1)=h_1(y_1) \leq h_1(y^*) \leq h(y^*).$$
    We notice that the first equality is due to $y_1\leq c$ and $c<d$, the first inequality is due to the optimality of $y_1$, and the second one to the fact that $h(y)\geq h_1(y)$, $\forall y$. We thus infer that, for this case, $h(y^*)\geq h(y_1)$, implying $y^*=y_1$. 
    Suppose now that $y_2\in[c,d]$. Under this assumption, we obtain analogously $h(y_2)=h_2(y_2) \leq h_2(y^*) \leq h(y^*)$,
    which implies $y^*=y_2$. The assumption $y_3\geq d$ translates into $h(y_3)=h_3(y_3) \leq h_3(y^*) \leq h(y^*)$,
    that is, $y^*=y_3$. 
    In order to finish the proof it suffices to check when the optimum is attained at the non differentiable points $c$ and $d$. 
    For this purpose, we analyze the limiting derivatives at $c$ and $d$ respectively
    \begin{align*} \lim\limits_{y\to c^-} h'(y) &= c{-}b{-}\alpha =c{-}y_1, \, \lim\limits_{y\to c^+} h'(y) = c{-}b =c{-}y_2, \\
    \lim\limits_{y\to d^-} h'(y) &= d{-}b =d{-}y_2, \, \lim\limits_{y\to d^+} h'(y) = d{-}b {+} \alpha=d{-}y_3.
    \end{align*}
    From the previous expressions, it is simple to check that the conditions $y_1>c$ and $y_2<c$ imply that $0$ belongs to the sub-differential of $h(y)$ at $y=c$, leading to $y^*=c$. Similarly,~the conditions $y_2>d$ and $y_3<d$ imply that $0$ belongs to the sub-differential of $h(y)$ at $y=d$, which translates into $y^*=d$. \qed

\end{appendices}
	
\end{document}